\documentclass[12pt,preprint]{aastex}
\usepackage{natbib}

\newcommand{\mc}{\multicolumn}

\newcommand{\ngh}{\dot{N}_{\rm ghost}}
\newcommand{\nr}{N_{\rm relic}}
\newcommand{\ngal}{N_{\rm rgal}}
\newcommand{\rr}{radio relic }

\begin{document}

\title{The Luminosity Function of Cluster Radio Relics}

\author{M. Br\"{u}ggen\altaffilmark{1}, T.A. En{\ss}lin\altaffilmark{2}, F. Miniati\altaffilmark{2}}
\altaffiltext{1}{International University Bremen, Campus Ring 1, 28759 Bremen, Germany}
\altaffiltext{2}{Max-Planck Institut f\"ur Astrophysik,
Karl-Schwarzschild-Str 1, 85740 Garching, Germany} 

\email{m.brueggen@iu-bremen.de}

\begin{abstract}

In this paper we compute the luminosity function of radio relics. In
our calculation we include only those relics that are produced by the
compression of former radio cocoons. This compression is provided by
shocks that are generated in the process of structure
formation. Starting from an analytical model for the luminosity
evolution of ageing radio cocoons, the luminosity function of radio
galaxies and the statistics of shocks as inferred from cosmological
simulations, we are able to make the first estimates of the brightness
distribution of radio relics. The computed luminosity function is
consistent with current observations and predicts that more than
$10^3$ radio relics should be discovered with the upcoming generation
of low-frequency radio telescopes. Moreover, we predict that radio
relics are predominantly found in low-pressure regions outside the
cores of clusters.

\end{abstract}

\section{Introduction}

A number of diffuse, steep-spectrum radio sources without optical
identification has been observed in galaxy clusters. These sources
have complex morphologies and show diffuse and irregular emission
\citep{kempner:01,bacchi:03}. They are usually subdivided into two
classes, denoted as `radio halos' and `radio relics' (a more detailed
classification scheme is put forward by \citet{kempner:03}).  Cluster
radio halos are unpolarised and have diffuse morphologies that are
similar to those of the thermal X-ray emission of the cluster gas.
Unlike halos, radio relics are typically located near the periphery of
the cluster; they often exhibit sharp emission edges and many of them
show strong radio polarisation. For more details on observations of
diffuse cluster radio sources the reader is referred to
\citep{kempner:01, giovannini:99}.

The origin of these diffuse radio sources is still not entirely
clear. Two processes for the formation of radio relics have been
proposed, both of which invoke the action of shock waves. Shock waves
are produced in the course of structure formation, e.g. by mergers
between, and accretion onto clusters of galaxies \citep{miniati:00},
and may provide the necessary acceleration of the electrons. The two
most important mechanisms that are believed to produce radio relics
are: (i) in-situ diffusive shock acceleration by the Fermi~I process
\citep{ensslin:98,roettiger:99,miniati:01} and (ii) re-acceleration of
electrons by compression of existing cocoons of radio plasma
\citep{ensslin:01, ensslin:02}. In all the aforementioned scenarios
the diffuse radio emission is associated with shock-fronts. Some of
the relics are characterised by highly structured morphologies and
those are more likely to have been produced by the compression
scenario \citep{ensslin:02}. In this paper we will focus on the latter
types of relics, i.e. those that are produced by compression of old
radio cocoons.

When a radio ghost, an old invisible radio galaxy cocoon, is passed by
a cluster merger shock wave with a typical velocity of a few 1000 km/s
the ghost is compressed adiabatically and not shocked because of the
much higher sound speed within it. Therefore, diffusive shock
acceleration is unlikely to be the prime mechanism that re-energises
the relativistic electron population. However, it has been shown that
the energy gained during the adiabatic compression, together with the
increase in the magnetic fields strength, can cause the old radio
cocoon to emit radio waves again. \citet{ensslin:01} showed that the
spectra thus produced are consistent with an old electron population
that has been adiabatically compressed. With the aid of
magneto-hydrodynamical simulations \citet{ensslin:02} demonstrated
that the resulting relics typically possess a toroidal shape and are
partially polarised - in good agreement with observations.

In this paper we attempt to predict the radio-luminosity function of
cluster radio relics, in particular of radio relics that are produced
via compression of old radio plasma \footnote{Recently, a new
nomenclature has been proposed where this type of relic is called
'radio phoenix'.}. Our predictions are intended to provide an
order-of-magnitude estimate for the potential detection of diffuse
radio sources by future high-sensitivity radio observatories, such as
the LOw Frequency ARray (LOFAR) and the Square Kilometer Array. It is
too early to aim for great precision in this calculation. Here, we
merely attempt to devise a simple and physically intuitive model which
will give a first estimate of the luminosity function.  The paper is
organised as follows: In the following section we explain conceptually
our approach in calculating the radio relic luminosity function. In
Sec. III we describe the cosmological simulations that provide the
input for our calculations, which then are presented in detail in
Sec. IV. Finally, the results are presented and discussed in Sec. V.

\section{Outline}

The progenitors for radio relics are so-called 'radio ghosts', which
are radio-quiet bubbles of non-thermal plasma released by formerly
active radio galaxies. In the picture of relic formation considered
here, the 'ghost' bubbles are re-energised by compression by
large-scale shocks. Consequently, for a calculation of the relic
luminosity function, we have to determine the statistics of, both, the
relevant properties of radio ghosts and of cosmic shock waves.

\noindent 
{\bf Radio ghosts:}\\
In order to calculate the statistics of radio ghosts, we start from
the luminosity distribution of radio galaxies that form the
progenitors of radio ghosts. In the next step, the observed radio
galaxy luminosity function needs to be translated into a birth rate of
radio ghosts, and this is done by assuming a constant lifetime for the
radio galaxy of $t_{\rm bubble}\sim 10^7$ yrs. This means that the
radio cocoon is observable on average for $\sim 10^7$ yrs.

This step is a coarse simplification since it ignores any intrinsic
luminosity evolution during the observable lifetime of a radio galaxy
\citep{kaiser:02,blundell:00,kaiser:97,kaiserdennett:97}. Its
justification lies partially in the fact that a more sophisticated
treatment would require the introduction of several poorly constrained
parameters, an expense which is not rewarded by the only moderate gain
in accuracy. The observed radio luminosity function is dominated by
the long-lasting late stage in the evolution of radio galaxies in
which the radio luminosity is nearly constant. Therefore, the
approximation that each source only has a single luminosity during its
lifetime is acceptable compared to other uncertainties.

Now, the luminosity of a radio ghost is boosted when it is
compressed by a shock wave. Whether this leads to an observable \rr
depends on a combination of shock strength and age of the radio
ghost. The shock has to be strong enough to lead to a sufficient
amplification of the magnetic field inside the ghost and to a
sufficient increase of the energies of the most relativistic
electrons.  Only a strong enough shock allows the renewed emission of
synchrotron emission at observable frequencies. However, the maximal
energy in the electron spectrum decreases with age owing to the
unavoidable synchrotron and CMB inverse-Compton energy losses. This
implies that the radio ghost must not exceed a certain age, if it is
to be re-activated by a shock wave of a given strength, and,
therefore, we need to estimate the age distribution of radio ghosts.

The birth rate of radio ghosts changes drastically during cosmological
times, as we know from the strongly redshift-dependent radio
luminosity function. However, on the timescale of $\sim 10^8$ yrs,
over which radio ghosts are able to be re-activated by typical shock
waves, the radio luminosity function is roughly constant. Therefore,
instead of modelling the demographic evolution of the ghost
distribution, for simplicity we use a flat age distribution, even
though \citet{kaiser:99} do not fully confirm this approximation. This
implies that the number of ghosts is given by

\begin{equation}
\label{eq:demography1}
\dot{N}_{\rm ghost} = \frac{N_{\rm rgal}}{t_{\rm bubble}}.
\end{equation}

In the next step, the maximal age, within which a radio ghost can be
re-activated, is calculated.  Because of the very low density inside
the radio cocoon, inverse Compton and synchrotron emission are the
dominant processes whereby the electrons lose their energy. Thus, the
maximal Lorentz factor of the relativistic electron population evolves
according to

\begin{equation}
\label{eq:cool1}
  \frac{d\gamma_{\rm max}}{dt} = - a_0\,(\epsilon_{B} +
  \epsilon_{\rm CMB})\,  \gamma_{\rm max}^2\, ,
\end{equation}

where $\epsilon_B$ and $\epsilon_{\rm CMB}$ are the magnetic field and
CMB energy densities respectively, and $a_0 = \frac{4}{3}\,
\sigma_{\rm T}/(m_{\rm e}\,c)$. This equation has the solution

\begin{equation}
\label{eq:cool2}
  \gamma_{\rm max}(t) = \left(\frac{1}{\gamma_{\rm max}(t_0)} +
  a_0\,(\epsilon_{B} + \epsilon_{\rm CMB})\,(t-t_0)\right)^{-1} \ ,
\end{equation}

for a period $t_0$ to $t$ with constant ambient conditions
($\epsilon_B$, $\epsilon_{\rm CMB} = $ const).  A shock wave of Mach
number $M$ compresses radio plasma with a relativistic equation of
state (adiabatic index of $\frac{4}{3}$) by a factor

\begin{equation}
  C = C(M) = \left (\frac{5M^2-1}{4}\right )^{3/4}
\end{equation}

and thereby increases the maximum electron energy adiabatically to
$\gamma_{\rm max,2} = \gamma_{\rm max,1} C^{\frac{1}{3}}$ and the
magnetic field energy density to $\epsilon_{B_2} = \epsilon_{B_1}
C^{\frac{4}{3}}$, where the index $1$ and $2$ labels pre- and
post-shock quantities respectively. For simplicity, radiative energy
losses during the short compression phase are neglected. The radio
ghost can be regarded to be reactivated, and therefore to form a \rr,
if the characteristic synchrotron frequency of the most energetic
electrons

\begin{equation}
  \nu_{\rm max,2} = \frac{3 \,e \, B_2}{2\,\pi\,m_{\rm e}\,c}\,
  \gamma_{\rm max,2}^2
\end{equation}

is above the observing frequency $\nu$. The relic remains visible for a
period $t_2$ until (the now enhanced) radiative cooling according to
Eqs. \ref{eq:cool1} and \ref{eq:cool2} has moved the synchrotron
cut-off frequency below $\nu$. 

Clearly, in order to compute the synchrotron ageing, we need a
description of the magnetic field strength inside the radio
ghosts. Here, we assume that the magnetic energy density is some
fraction of the total pressure inside the ghost, which itself is in
pressure equilibrium with the ambient pressure $p$. Hence, we need to
specify the distribution of radio ghosts as a function of the ambient
pressure, which we approximate as follows:

Our starting point is the spatial statistics of the radio galaxy
 distribution, which we scale with some power $\alpha$ of the ambient
 gas density $\rho$. We assume that the radio galaxy distribution
 $N_{\rm rgal}(L,\vec{r},z)$ per luminosity and for a specific volume
 element at location $\vec{r}$ can be written as a direct product of
 its global luminosity function $N_{\rm rgal}(L,z)$ and a term
 proportional to $\rho^\alpha$ (for a discussion see
 \citet{kaiser:99}). This allows us to construct the radio luminosity
 function as the number of radio galaxies of luminosity $L=L(\nu)$ per
 unit volume $V$, pressure $p$ and luminosity as a function of $p$,
 $L$ and redshift $z$

\begin{equation}
\ngal(L,p,z) \ dp \ dV \ dL = {\cal D}(L,z) R_{\alpha} (p) \ dp \ dV
\ dL\ .
\end{equation}
One can get a handle on $\alpha$ by demanding that $\sim 30$ \% - 50
\% of all radio galaxies lie in clusters
\citep{ledlow:95,ledlow:96}. If we characterise clusters by means of
their pressures, i.e. as environments where the pressure lies above
$\sim 10^{-12}$ erg cm$^{-3}$, we find that for $\alpha$ of around 1.5
about 30 \% of radiogalaxies should lie in clusters and for $\alpha =
2$ about 60 \% would lie in clusters. One should note, however, that
these estimates are very crude and should be treated with caution. So
in the following, we will present our results for these two values of
$\alpha$.

The distribution function $R_{\alpha} (p)$ is defined via

\begin{equation}
R_{\alpha} (p') dp'= \frac{A}{V} dp'\int dV\ \rho({\bf r})^{\alpha}
\delta(p({\bf r})-p') \ ,
\end{equation}
where $\delta$ is the delta function and $A$ is a normalisation
constant chosen such that $\int dp\ R_{\alpha} (p) = 1$.

As explained in more detail later, $R_{\alpha} (p)$ is extracted from
a simulation of cosmological structure formation. Now we can also
express the birth rate of radio {\it ghosts} per unit $p$, $L$ and $V$

\begin{equation}
\ngh\ dp\ dV\ dL = \frac{\ngal}{t_{\rm bubble}} \ dp\ dV\ dL .
\end{equation}

\noindent 
{\bf Shock waves:}\\
When the radio ghost is compressed adiabatically, its synchrotron
luminosity, $L_2$, increases with respect to its initial luminosity,
$L_1$, by a factor $L_2/L_1= [(5M^2-1)/4]^{5/2}$
\citep{ensslin:01}. This treatment ignores subtle spectral steepening
effects that occur when the spectral cut-off passes the observing
frequency\footnote{ The aging of the relativistic electron
populations, which was originally a power-law in Lorentz-factor
$\gamma$ with a spectral index $\alpha_{\rm e}$
\begin{displaymath}
  f(\gamma,t_1=0)\, d\gamma = f_0\,\gamma^{-\alpha_{\rm e}}\, d\gamma\,,
\end{displaymath}
is described by 
\begin{displaymath}
  f(\gamma,t_1)\, d\gamma = f_0\,\gamma^{-\alpha_{\rm
  e}}\,(1-\gamma/\gamma_{\rm max}(t_1))^{\alpha_{\rm e}-2}
  \theta(\gamma_{\rm max} -\gamma)\, d\gamma\,
\end{displaymath}
which shows a spectral steepening shortly below the cutoff for
$\alpha_{\rm e} >2$.}, and the detailed shape of the synchrotron
emission spectral kernel. However, our simplified treatment should be
entirely sufficient for our purposes.

Finally, we need to determine the frequency with which shock waves of
a given Mach number $M$ sweep over a radio ghost, which is situated at
redshift $z$ in an environment of pressure $p$. This frequency is
written as

\begin{equation}
\omega(p,M,z)\ dM= \frac{\dot{m}^{\alpha}(p,M,z)}{R_{\alpha}}\ dM \ ,
\end{equation}

where $\dot{m}^\alpha$ is the flow of $\rho^\alpha$ (a density-biased
mass flow) integrated over shocks of Mach number $M$, at redshift $z$
and in an environment of pressure $p$, given by

\begin{equation}
\dot{m}^{\alpha}(p,M,z) = \frac{1}{V}\int d{\bf A}_{\rm shock}\cdot {\bf v}_{\rm shock}\
\rho({\bf r})^{\alpha}\ \delta(M-v_{\rm shock}/c_s)\delta(p-\rho c_s^2/\gamma_{\rm
gas})\ ,
\end{equation}
where $V$ is the volume considered, $c_s$ the sound speed,
${\bf v}_{\rm shock}$ the shock velocity and $\gamma_{\rm
gas}=\frac{5}{3}$ the adiabatic index of the gas. Note that $p,\rho$
and $c_s$ are evaluated ahead of the shock.

It is useful to define the time between successive shocks of any
strength, which is given by

\begin{equation}
\tau (p,z) = \left [\int_0^{\infty} dM\ \omega(p,M,z)\right ]^{-1}. \label{tau}
\end{equation}

From the simulations that will be described in the following section,
this time was inferred to be of the order of 3 Hubble times, which is
roughly consistent with the findings by \citep{cen:99}, who claim that
about 30 \% of the baryons in the universe at redshift zero have been
shocked. We also find that $\tau$ increases with pressure which is
attributed to the fact that there are many more low-pressure regions
in the universe and that these regions have lower sound speeds and
shock more easily.

\section{Numerical Simulation of the Large Scale Structure}

The formation and evolution of the large-scale structure is computed
by means of an Eulerian, grid based Total-Variation-Diminishing
hydro+N-body code \citep{ryu:93}.

We adopt a canonical, flat $\Lambda$CDM cosmological model with a
total mass density $\Omega_m=0.3$ and a vacuum energy density
$\Omega_\Lambda= 1- \Omega_m= 0.7$.  We assume a normalized Hubble
constant $h\equiv H_0/100$ km s$^{-1}$ Mpc$^{-1}$ = 0.67 and a
baryonic mass density, $\Omega_b=0.04$.  The simulation is started at
redshift $z\simeq 60$ with initial density perturbations generated as
a Gaussian random field and characterized by a power spectrum with a
spectral index $n_s=1$ and ``cluster-normalization'' $\sigma_8=0.9$.
We adopt a computational box size of $50\,h^{-1}\,$Mpc. In this box
the dark matter component is described by 256$^3$ particles whereas
the gas component is evolved on a comoving grid of 512$^3$ zones. Thus
each numerical cell measures about $100\,h^{-1}\,$kpc (comoving) and
each dark matter particle corresponds to $2\times
10^9\,h^{-1}\,M_\odot$.  Further details on the cosmological
simulations can be found in \citet{miniati:02}.

The results from the simulation are used to build the distribution
functions $\dot{m}_\alpha$ and $R_\alpha$ that are required in our
calculations.  As in previous studies \citep{miniati:00}, we identify
shocks as converging flows (${\bf \nabla \cdot v < 0}$) that
experience a pressure jump $\Delta P/P$ above a threshold
corresponding to a Mach number M=1.5.  Once a shock has been
identified the jump conditions are evaluated from the numerical
solution. Finally, the shock speed and Mach number are computed
\citet{landau:87}.

\section{Calculations}


In this section we carry out the calculations that were outlined in
prose in the previous sections. The total number of radio {\it relics}
of luminosity $L_2$ per unit volume at a given redshift is given by
the birth rate of radio {\it ghosts}, $\ngh$, times the time for which
the radio relic is visible, $t_2$, times the frequency of shocks of
given strength, $\omega$ that can boost their luminosity from $L_1$ to
$L_2$. All this is integrated over the luminosity of the {\it ghosts},
$L_1$, the age of the {\it ghost} at the time of compression, $t_1$,
and pressure and Mach number, i.e.

\begin{eqnarray}
\nr(L_2,z) & = & \int dt_1 \int dL_1 \int dp \int dM\ \ngh(L_1,p,z)
t_2(t_1,p,M,z) \nonumber \\
& & \omega(p,M,z) e^{-t_1/\tau} \delta [L_2 -
((5M^2-1)/4)^{5/2} L_1] ,
\end{eqnarray}
where $\tau$ is given by Eq.(\ref{tau}) and $t_2(t_1,p,M,z)$ the time
for which the relic is visible, i.e., until the cut-off frequency
falls below the observing frequency. This time is calculated
below. Note that we have introduced a factor of $e^{-t_1/\tau}$ which
ensures that a radio ghost can only be reactivated once.

We can perform the integration with respect to $L_1$, by assuming that,
upon compression, the luminosity of the radio ghosts increases by

\begin{equation}
\frac{L_2}{L_1} = \left ( \frac{p_2}{p_1} \right )^{5/2} = \left (
\frac{5M^2-1}{4} \right )^{5/2} =: g(M),
\end{equation}
where $L_2$ and $L_1$ and $p_2$ and $p_1$ refer to the post-/pre-shock
luminosities and pressures, respectively. Moreover, in concordance
with observations, a spectral index of 1 for the synchrotron emission
was assumed \citep{ensslin:02}.  Note that we make the simplifying
assumption that the ghost instantaneously flares up to $L_2$ as the
shock sweeps by. Performing the $L_1$-integration, we obtain

\begin{equation}
\nr(L_2,z) = \int dp \int dM\ \ngh(L_2/g(M),p,z)\frac{\omega(p,M,z)}{g(M)} \int dt_1\ t_2(t_1,p,M,z) e^{-t_1/\tau}
\ . \label{NREL}
\end{equation}

The 'visibility time' $t_2$ is calculated as follows: The cut-off
frequency, $\nu_c$, decreases due to synchrotron radiation and Compton
scattering with the CMB photons. Furthermore, the compression of the
radio ghost after a time $t_1$ boosts the cut-off frequency by a
factor of $(p_2/p_1)^{1/2}$. Hence, $\nu_c$ is given by

\begin{equation}
\nu_c = \frac{3eB_2}{2\pi m_e c a_0^2}\left [ \left (
\frac{p_2}{p_1}\right )^{-1/4}(\epsilon_{B_1} + \epsilon_{\rm
CMB})t_1+(\epsilon_{B_2} + \epsilon_{\rm CMB})t_2\right ]^{-2} \ ,
\end{equation}
where $B_2 = B_1 (p_2/p_1)^{1/2}$ is the magnetic field strength in
the ghost after compression (note that $B_1=\sqrt{\eta_{\rm B} 8\pi p_1}$),
$\epsilon_{\rm CMB}$ the energy density of the CMB ($4.02\times
10^{-13}$ erg cm$^{-3}$ at $z=0$) and $a_0=\frac{4\sigma_{\rm T}}{3m_e
c}$. Here, we assumed that the magnetic energy density is some
fraction of the total pressure in the ghost $\eta_{\rm B} \approx 0.3$.

Solving for $t_2$ yields:

\begin{equation}
t_2= \left [\left (\frac{3eB_2}{2\pi m_e c a_0^2\nu}\right )^{1/2} - \left (
\frac{p_2}{p_1} \right )^{-1/4} (\epsilon_{B_1} + \epsilon_{\rm CMB})
t_1 \right ]/(\epsilon_{B_2} + \epsilon_{\rm CMB}) \ .
\end{equation}

Now we calculate the last integral in Eq. (\ref{NREL}), i.e. $I = \int
dt_1\ t_2 \exp(-t_1/\tau)$, where it is convenient to introduce the
definitions for the maximal 'life time' of the relic:

\begin{equation}
t_{\rm max} = \frac{1}{\epsilon_{B_2} + \epsilon_{\rm
CMB}}\left (\frac{3eB_1}{2\pi m_e ca_o^2\nu} \right )^{1/2}\left (
\frac{5M^2-1}{4}\right )^{1/2} \ ,
\end{equation}
and the ratio of the loss rates before and after compression:

\begin{equation}
r = \frac{\epsilon_{B_1} + \epsilon_{\rm CMB}}{\epsilon_{B_2} +
\epsilon_{\rm CMB}} \ .
\end{equation}

Now the integral, $I$, reduces to:

\begin{equation}
I = (p_2/p_1)^{-1/4} \int_0^{t_{\rm max}/r} dt_1\ (t_{\rm max} - r
t_1) \exp(-t_1/\tau) \ ,
\end{equation}

which can be written as

\begin{equation}
I = (p_2/p_1)^{-1/4} [r\tau t_{\rm max} e^{-t_{\rm max}/r\tau}+(\tau t_{\rm
max}-r\tau^2)(1-e^{-t_{\rm max}/r\tau}) ]
\end{equation}
or

\begin{equation}
I = \left (\frac{5M^2-1}{4}\right )^{-1/4} [\tau t_{\rm max} -r\tau^2(1-e^{-t_{\rm max}/r\tau}) ]\ ,
\end{equation}
where $r$ and $t_{\rm max}$ are functions of $p,M,z$ and $\tau$ a
function of $p$ and $z$.

Hence the number of radio relics per unit volume is given by the integral:

\begin{equation}
\nr(L_2,z) = \int dp \int dM\ \ngh(L_2/g(M),p,z) I(p,M,z) \omega(p,M,z)/g(M) 
\end{equation}

\begin{eqnarray}
\nr(L_2,z) & = & \int dp \int dM\ \ngh(L_2/g(M),p,z) \left
(\frac{5M^2-1}{4}\right )^{-11/4} \omega \tau \nonumber \\
& & [t_{\rm max} - r\tau(1-e^{-t_{\rm max}/r\tau}) ] \  .
\label{eq:nr}
\end{eqnarray}

For the radio luminosity function we used 'model A' from \citet{willott:01}:

\begin{equation}
{\cal D}(L,z) =\rho_{\rm l} +\rho_{\rm h} \ ,
\end{equation}
where
\begin{equation}
\rho_{\rm l} =\rho_{{\rm l}\circ} \left( \frac{L}{L_{{\rm l} \star}} \right) ^{-\alpha_{\rm l}} \exp  {\left( \frac{-L}{L_{{\rm l} \star}} \right) } (1+z)^{k_{\rm l}}  ~~~~~ {\rm for}~ z<z_{{\rm l}\circ},
\end{equation}
\begin{equation}
\rho_{\rm l} =\rho_{{\rm l}\circ} \left( \frac{L}{L_{{\rm l} \star}} \right) ^{-\alpha_{\rm l}} \exp  {\left( \frac{-L}{L_{{\rm l} \star}} \right) } (1+z_{{\rm l}\circ})^{k_{\rm l}} ~~~ {\rm for}~ z\geq z_{{\rm l}\circ},
\end{equation}
\begin{equation}
\rho_{\rm h} =\rho_{{\rm h}\circ} \left( \frac{L}{L_{{\rm h} \star}} \right) ^{-\alpha_{\rm h}} \exp  {\left( \frac{-L_{{\rm h} \star}}{L} \right) } f_{\rm h} (z). 
\end{equation}

The high--luminosity evolution function  $f_{\rm h} (z)$ takes the form:
\begin{equation}
f_{\rm h}(z) = \exp \left\{ - \frac {1}{2} {\left( \frac{z-z_{{\rm
h}\circ}}{z_{{\rm h}1}} \right)^{2} } \right\} \ ,
\end{equation}
where the different parameters for this model are given in Table
1. Finally, the remaining two integrals over $p$ and $M$ in
Eq. \ref{eq:nr} will have to be done numerically.

\section{Results and Discussion}

\begin{figure}[htp]
\plotone{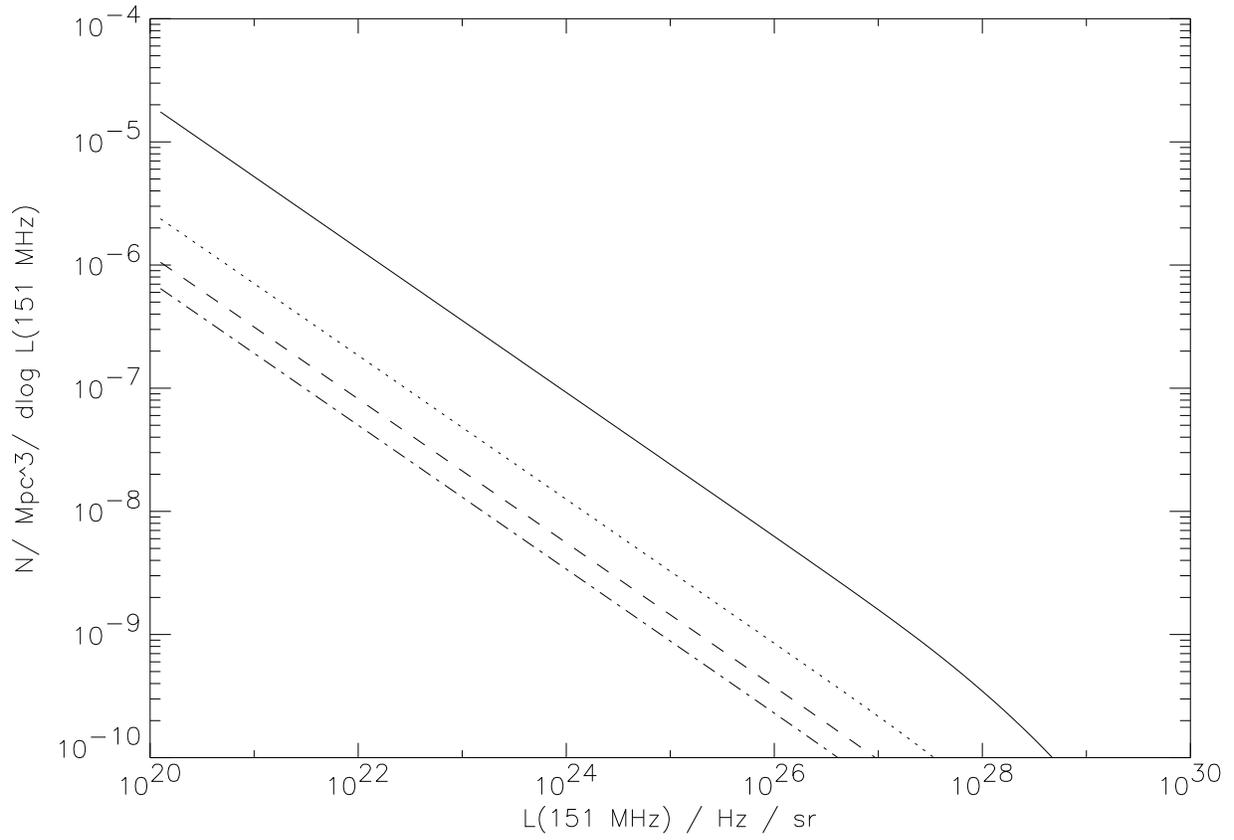}
\caption{Radio relic luminosity function at 151 MHz for $\alpha = 2$ at $z=0.001$
(solid), 0.75 (dotted),  1.5 (dashed) and 3 (dot-dashed).}
\label{fig1}
\end{figure}

The number density of relics per decade in luminosity is shown in
Fig. \ref{fig1}.  Between $L=10^{20}$ and $10^{29}$ W Hz$^{-1}$
sr$^{-1}$ the number or radio relics per unit volume follows a power
law $N\propto L^{\beta}$ with $\beta = -0.6$. As indicated in the plot
the number density of relics goes down with increasing redshift.

\begin{figure}[htp]
\plotone{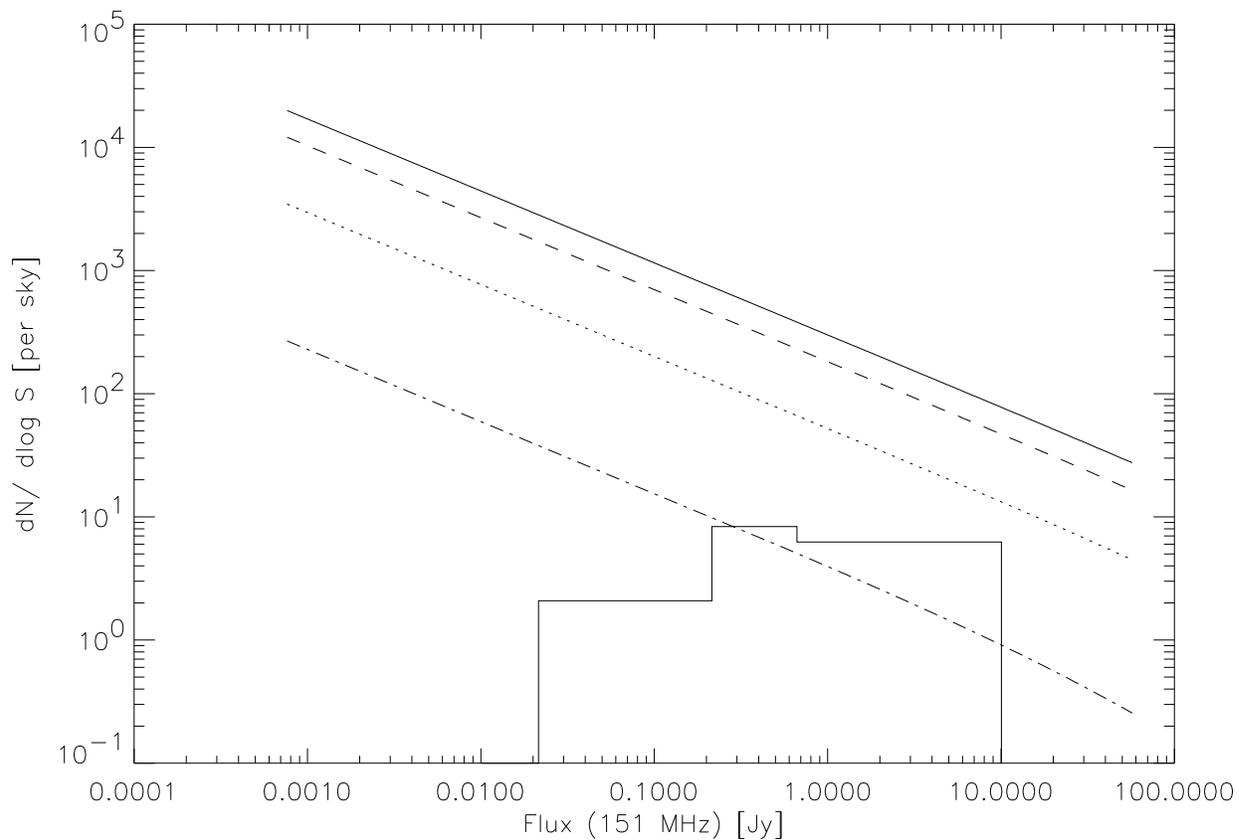}
\caption{Expected brightness distribution for radio halos at
151 MHz
(assumed $h=0.7$, $\Omega_m=0.3$). The solid line corresponds to the
total number of relics, while the dashed line includes only the
contribution from relics that are
found at pressures $p>10^{-14}$ g cm$^{-3}$, the dotted line to those at
$p>10^{-13}$ g cm$^{-3}$, and the dot-dashed line only those at
$p>10^{-12}$ g cm$^{-3}$. Here $\alpha$ was taken as 1.5.}
\label{fig2}
\end{figure}

\begin{figure}[htp]
\plotone{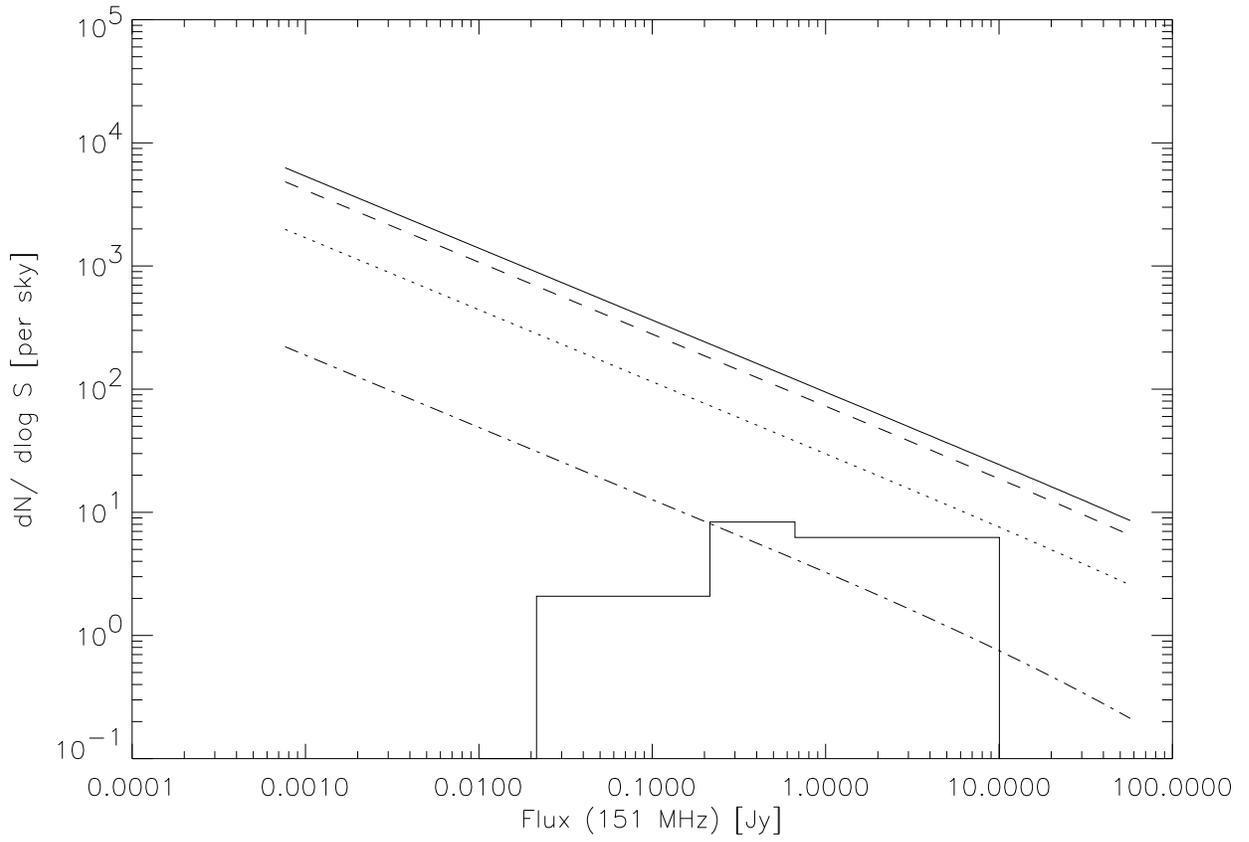}
\caption{Same as Fig. \ref{fig2} except that here $\alpha$ was taken as 2.}
\label{fig3}
\end{figure}

The number of relics per decade in brightness at 150 MHz as a function
of flux is shown in Figs. \ref{fig2} and \ref{fig3} for $\alpha=1.5$
and 2, respectively. Their distribution roughly follows a power law of
index -0.6. For higher values of $\alpha$ the number of relics
decreases because, for higher $\alpha$, a larger fraction of radio
ghosts is concentrated in the cores of clusters, where the lifetime of
the radio ghosts is shorter and the chances for a reappearance as a
radio relics diminish. The number of radio relics at 151 MHz is higher
than the number of {\it radio halos} at 1.4 GHz by a factor of $\sim
10$ as predicted by \citet{ensslin3:02}. Assuming a spectral index of
1, as we did in this paper, this would imply that the numbers of radio
relics and halos should roughly match if compared at the same
frequency, which is what is indeed found. In Fig. \ref{fig2} we have
included a histogram with the observed flux density distribution of
the sample of Feretti (1999). Even though the current sample of radio
relics is still poor, we find that current observations are not
inconsistent with our predictions. Ignoring the uncertainties of our
model, one would conclude from Figs. \ref{fig2} and \ref{fig3} that,
in the range between 1 and 10 Jansky, approximately 10 \% of the radio
relics have been detected.  However, we fear that the predicted number
may be a bit high and that we have overestimated the luminosities due
to two reasons: First, we neglected the enhanced cooling during the
compression of the bubble and, second, the majority of radio cocoons
will not be in exact pressure equilibrium but are more likely to be
still somewhat overpressured when they are being compressed. These two
effects will shift the curves to the left while leaving the slope
unchanged.

In order to determine in which environments the majority of relics is
to be found, we computed the contribution to the total flux density
from relics that lie within a certain pressure range (since pressure
is the free variable in our formalism). The solid line in
Fig. \ref{fig2} corresponds to the total number of relics, while the
dashed line includes only the contribution from relics that are found
at pressures $p>10^{-14}$ g cm$^{-3}$, the dotted line to those at
$p>10^{-13}$ g cm$^{-3}$, and the dot-dashed line only those at
$p>10^{-12}$ g cm$^{-3}$. We find that the majority of radio relics is
found at pressures between a bit below $10^{-14}$ g cm$^{-3}$ and
$10^{-13}$ g cm$^{-3}$, which are typical values for pressures at the
edges of galaxy cluster. This is in good agreement with the observed
locations of radio relics: nearly all of them have been found in the
periphery of galaxy clusters. However, similar pressure conditions are
also encountered around groups and in filaments of galaxies, where we
would also expect that future high-sensitivity observations detect
radio relics. Conversely, in the cores of clusters, i.e. at typical
pressures of $p<10^{-12}$ g cm$^{-3}$, only $\sim$ 1\% of radio relics
are expected (see dot-dashed line). The fact, that shocks become
stronger when they leave the cluster, more than makes up for the fact
that there are fewer radio galaxies in the outer regions. Moreover,
radio plasma is very short-lived at high ambient pressures because the
synchrotron losses are high. Thus, even though there is more radio
plasma in the cores of clusters, only little of it is fresh enough to
form radio relics again. Here we predict that radio relics are
predominantly found in low-pressure regions outside the cores of
clusters.

Using very simple assumptions we were able to give a first gross
estimate of the number of radio relics that are expected at different
flux limits. Upcoming low-frequency radio telescopes such as LOFAR are
ideally suited to search for relics at the edges of clusters. LOFAR is
expected to have a point source sensitivity of 0.13 mJy at 120 MHz
within one hour of integration time and a 4 MHz bandwidth. A survey
covering half the sky can be accomplished in a years timescale at this
frequency and with this depth \citep{ensslin3:02}. Such a survey is
likely to find of the order of $10^4$ radio relics. As mentioned
earlier, we believe that this number may be a bit too high because the
number of detected relics between 1 and 10 Jansky is lower by a factor
of roughly 10. However, given the simplicity of our model and the
scarceness of data, this is not a bad match. More importantly, while
the normalisation of the brightness distribution is sensitive to some
of our assumptions and, consequently, somewhat uncertain, both, the
slope of the distribution and the relative contributions from
different pressure bins are fairly robust predictions.

Finally, let us summarise the assumptions that went into the
calculation of the radio relics luminosity function:

(i) Radio relics are produced by the adiabatic compression of fossil radio
plasma (radio ghosts).

(ii) The birth rate of radio ghosts is given by the number of radio
galaxies divided by a constant activity time span, $t_{\rm bubble}$.

(iii) The number density of radio galaxies is given by the radio
luminosity function \citep{willott:01} scaled with a power of the
local density. This power was assumed to be either 1.5 or 2.

(iv) The compression of the radio plasma is instantaneous upon passage
of a shock.

(v) No stochastic particle reacceleration takes place and magnetic
fields are merely passive (magnetic energy density = 0.3 of total
pressure).

(vi) Within the time within which the radio ghosts can be reactivated,
they do not migrate (by buoyancy or advection).

\acknowledgments We thank M. Hoeft for valuable discussions. MB thanks
the Max-Planck Insitut f\"ur Astrophysik in Garching for their kind
hospitality. FM acknowledges support by the Research and Training
Network ``The Physics of the Intergalactic Medium'' under EU contract
HPRN-CT2000-00126 RG29185.

\begin{table*}[htp]
\begin{center}
\begin{tabular}{ccccccccccc}
\hline\hline 
\mc{1}{c}{Model} &\mc{1}{c}{$\log(\rho_
{{\rm l}\circ})$}&\mc{1}{c}{$\alpha_{\rm l}$}&\mc{1}{c}{$\log(L_{{\rm
l} \star})$}&\mc{1}{c}{$z_{{\rm l}\circ}$}&\mc{1}{c}{$k_{\rm
l}$}&\mc{1}{c}{$\log$ ($\rho_ {{\rm h}\circ}$)}&\mc{1}{c}{$\alpha_{\rm
h}$}&\mc{1}{c}{$\log(L_{{\rm h} \star})$}&\mc{1}{c}{$z_{{\rm
h}\circ}$}&\mc{1}{c}{$z_{{\rm h}1}$} \\
\hline\hline

A & $-7.503$ & $0.584$ & $26.46$ & $0.710$ & $3.60$ 
  & $-6.740$ & $2.42$  & $27.42$ & $2.23$  & $0.642$ \\

\hline\hline         
\end{tabular}
\end{center}              
\caption{Parameters chosen for the radio luminosity function (Model A from \citet{willott:01}).}
\end{table*}

\begin{table*}[htp]
\begin{center}
\begin{tabular}{ccccccc}
\hline \mc{1}{c}{Cluster} &\mc{1}{c}{z} &
\mc{1}{c}{P$_{1.4}$[$10^{23}$ W/Hz]} & \mc{1}{c}{L.S.[Mpc]} &
\mc{1}{c}{L$_{Xbol}$[$10^{44}$ erg/s]} & \mc{1}{c}{T[keV]} &
\mc{1}{c}{d [Mpc]} \\ \hline\hline

A85   &      0.0555 &  6.26  & 0.48 &  19.52 &   5.1  & 0.54   \\
A115  &      0.1971 &  255.5 & 1.88 &  31.09 &   4.9  & 0.93   \\
A610  &      0.0956 &  7.65  & 0.57 &  -     &    -   &  0.71  \\
A1300 &      0.3071 &  92.4  & 0.95 &  47.63 &   10.5 & 0.80   \\
A1367 &      0.0216 &  0.71  & 0.29 &   2.87 &   3.5  & 0.83   \\
A1656 &      0.0232 &  7.03  & 1.17 &  20.42 &   8.2  & 2.72   \\
A2255 &      0.0809 &  3.51  & 0.98 &  12.42 &   5.4  & 1.23   \\
A2256 &      0.0581 &  4.48  & 1.11 &  18.39 &   7.5  & 0.59   \\
A2744 &      0.3080 &  74.3  & 1.84 &  62.44 &   11.0 & 1.91   \\
A3667 &      0.0552 &  323.1 & 2.63 &  22.70 &   7    & 2.45   \\

\hline\hline         
\end{tabular}
\end{center}
\caption{Col. 1: cluster name; Col. 2: cluster redshift; Col. 3: radio
power of the diffuse source at 1.4 GHz; Col. 4: largest linear size of
the diffuse source; Col. 5: cluster X-ray bolometric luminosity; Col
6: cluster temperature obtained by averaging values in the literature;
Col. 7: projected distance of the diffuse source from the cluster
centre (from \citet{feretti:99})}
\end{table*}

\newpage


\end{document}